\def\sorb{b}
\if s\sorb
  \documentstyle{article}
  \newlength{\absize}
  \renewcommand{\baselinestretch}{1.5}
  \renewcommand{\arraystretch}{1.5}
  
\begin{document}
  \date{}
  \pagestyle{empty}
  \thispagestyle{empty}
  \renewcommand{\thefootnote}{\fnsymbol{footnote}}
  \newcommand{\starttext}{\newpage\normalsize
    \pagestyle{plain}
    \setlength{\baselineskip}{4ex}\par
    \twocolumn\setcounter{footnote}{0}
    \renewcommand{\thefootnote}{\arabic{footnote}}}
\else
  \documentstyle[12pt,a4wide,epsf]{article}
  \newlength{\absize}
  \setlength{\absize}{\textwidth}
  \renewcommand{\baselinestretch}{2.0}
  \renewcommand{\arraystretch}{2.0}
  \begin{document}
  \thispagestyle{empty}
  \pagestyle{empty}
  \renewcommand{\thefootnote}{\fnsymbol{footnote}}
  \newcommand{\starttext}{\newpage\normalsize
    \pagestyle{plain}
    \setlength{\baselineskip}{4ex}\par
    \setcounter{footnote}{0}
    \renewcommand{\thefootnote}{\arabic{footnote}}}
\fi
\newcommand{\preprint}[1]{%
  \begin{flushright}
    \setlength{\baselineskip}{3ex} #1
  \end{flushright}}
\renewcommand{\title}[1]{%
  \begin{center}
    \LARGE #1
  \end{center}\par}
\renewcommand{\author}[1]{%
  \vspace{2ex}
  {\Large
   \begin{center}
     \setlength{\baselineskip}{3ex} #1 \par
   \end{center}}}
\renewcommand{\thanks}[1]{\footnote{#1}}
\renewcommand{\abstract}[1]{%
  \vspace{2ex}
  \normalsize
  \begin{center}
    \centerline{\bf Abstract}\par
    \vspace{2ex}
    \parbox{\absize}{#1\setlength{\baselineskip}{2.5ex}\par}
  \end{center}}

\setlength{\parindent}{3em}
\setlength{\footnotesep}{.6\baselineskip}
\newcommand{\myfoot}[1]{%
  \footnote{\setlength{\baselineskip}{.75\baselineskip}#1}}
\renewcommand{\thepage}{\arabic{page}}
\setcounter{bottomnumber}{2}
\setcounter{topnumber}{3}
\setcounter{totalnumber}{4}
\newcommand{\figsize}{}
\renewcommand{\bottomfraction}{1}
\renewcommand{\topfraction}{1}
\renewcommand{\textfraction}{0}
\newcommand{\beq}{\begin{equation}}
\newcommand{\eeq}{\end{equation}}
\newcommand{\beqa}{\begin{eqnarray}}
\newcommand{\eeqa}{\end{eqnarray}}
\newcommand{\nn}{\nonumber}

\newcommand{\dd}{{\rm d}}
\newcommand{\mH}{m_{\rm H}}
\newcommand{\dLips}{{\rm dLips}}
%
\def\slash#1{#1 \hskip -0.5em /}
%
%

\preprint{University of Bergen, Department of Physics \\
Scientific/Technical Report No.\ 1993-03 \\ ISSN~0803-2696}

\vfill
\title{Angular and energy correlations in Higgs decay}

\vfill
\author{Arild Skjold \\ Per Osland \\\hfil\\
        Department of Physics\thanks{electronic mail addresses:
                {\tt \{skjold,osland\}@vsfys1.fi.uib.no}}\\
        University of Bergen \\ All\'egt.~55, N-5007 Bergen, Norway }
\date{}

\vfill
\abstract{We discuss how correlations among momenta of the
decay products may yield information about the
CP-parity of the Higgs particle.
These are correlations of decay planes defined by the momenta of pairs of
particles as well as correlations between energy differences.
Our study includes finite-width effects. In the narrow-width approximation
the angular correlations coincide with previously reported results.
The correlations between energy differences turn
out to be a much better probe for CP determination than the previously
suggested angular correlations, especially for massive Higgs bosons.
}

\vfill

\starttext


When some candidate for the Higgs particle is discovered, it becomes
imperative to establish its properties, other than the mass.
While the standard model Higgs boson is even under CP,
alternative and more general models contain Higgs bosons
for which this is not the case.
In specific models, production cross sections and branching ratios
will be different
\cite{HHG}
but it would clearly be valuable if one could establish its CP
on more general grounds, independent of the production mechanism.
We shall here discuss how decay distributions may shed light on
its intrinsic parity, in particular, how one can determine whether
it is even or odd under CP.

This situation is similar to the classical one of determining the parity of
the $\pi^0$ from the angular correlation of the planes spanned by the
momenta of the two Dalitz pairs \cite{Yang50,Dalitz},
\begin{equation}
\pi^0\rightarrow\gamma\gamma\rightarrow(e^+e^-)(e^+e^-).
\end{equation}

In that case, as pointed out by Yang
\cite{Yang50},
the correlation of the planes is unambiguously determined by
the parity-conserving $\pi^0\gamma\gamma$ and $\gamma e^+e^-$ couplings.
Classically, the $\pi^0\gamma\gamma$ coupling is proportional to
$\vec E\cdot\vec B$, where $\vec E$ and $\vec B$ are the electric and
magnetic field strengths, respectively.
In relativistic notation, this corresponds to the coupling
$i\epsilon^{\mu\nu\rho\sigma}F_{\mu\rho}F_{\nu\sigma}$,
where $F_{\mu\rho}$ is the electromagnetic field strength tensor.
If we describe the photons by the
vector field $A^\mu$, then,
with $p_1$ and $p_2$ the momenta of the two photons, this
$\pi^0\gamma\gamma$ coupling has the form \cite{Dalitz}
\begin{equation}
i \: g_{\rm eff}(p_1^2,p_2^2)\: \epsilon^{\mu\nu\rho\sigma}
p_{1\rho} p_{2\sigma},
\label{EQU:psuver}
\end{equation}
where $g_{\rm eff}$ is some effective coupling that depends on the kinematics.
For comparison, the coupling of the standard-model Higgs to vector bosons
(of mass $m_V$) is given by
\begin{equation}
i \: (2\,2^{1/4})\sqrt{G_{\rm F}}\,m_V^2g^{\mu\nu},
\label{EQU:stanguv}
\end{equation}
with $G_{\rm F}$ the Fermi constant.

In the electroweak interactions, parity is known to be violated.
However, we shall here assume that
parity violation is limited to the couplings between vector bosons
($W$ and $Z$) and fermions (quarks and leptons),
as is the case in the standard model.
Then it turns out that correlations similar to those of Dalitz pairs
in $\pi^0$ decay
determine the CP-parity of the Higgs boson.

In non-standard or extended models of the electroweak interactions,
there exist ``Higgs-like'' particles having negative CP.
An example of such a theory is
the minimal supersymmetric model (MSSM) \cite{MSSM}, where
there is a neutral CP-odd Higgs boson, often denoted $A^{0}$ and sometimes
referred to as a pseudoscalar.

We shall consider the decay where a standard-model Higgs (H) or
a `pseudoscalar' Higgs particle (A) decays via two
bosons ($W^+W^-$ or $ZZ$), to two
non-identical fermion-antifermion pairs,
\begin{eqnarray}
\label{EQU:Hdecay}
(H,A)\rightarrow V_1V_2\rightarrow (f_1\bar f_2)(f_3\bar f_4).
\end{eqnarray}
The two vector bosons need not be on mass shell.

If we let the momenta ($q_1$, $q_2$, $q_3$, and $q_4$)
of the two fermion-antifermion pairs
(in the Higgs rest frame) define two planes,
and denote by $\phi$ the angle between those two planes,
then we shall discuss the angular distribution of the decay rate
$\Gamma$,
\beq
\frac{1}{\Gamma}\:
\frac{{\mbox{{\rm d}}\Gamma}}{\mbox{{\rm d}}\phi}
\label{EQU:intro1}
\eeq
and a related quantity, to be defined below, both
in the case of CP-even and CP-odd Higgs bosons.

Related studies have been reported by \cite{Nel,Dun} in the context
of correlations between decay planes involving scalar and (technicolor)
pseudoscalar Higgs bosons.
It should be noted that the present discussion is more general,
since finite-width effects of the vector bosons are taken into account.
This enables us to investigate the angular correlations for Higgs masses below
the threshold for decay into real vector bosons.
In addition, we discuss correlations between energy differences. It
turns out that under suitable experimental conditions these correlations
provide an even better signal for CP determination.

We let $g_V$ and $g_A$ denote the vector and axial-vector parts
of the couplings, such that the fermion-fermion-vector coupling
is given by
\beq
- \frac{i}{2 \sqrt{2}} \gamma^\mu(g_V-g_A\gamma_5). \label{EQU:Dj90}
\eeq
In order to parametrize the vector and axial couplings, we define
the couplings involving vector bosons $V_1$ and $V_2$
in terms of angles $\chi_1$ and $\chi_2$ as
\beq
g^{(i)}_{V}  \equiv  g_i \cos \chi_i, \qquad
g^{(i)}_{A} \equiv g_i \sin \chi_i, \qquad i=1,2.
\label{EQU:Dj9}
\eeq
The only reference to these angles is through $\sin2\chi$.
Relevant values are given in table~1
\cite{Datagr}.
Furthermore, $g_{1}$ and $g_{2}$ contain CKM matrix elements and electric
and weak isospin charges relevant to the fermions in question.
The couplings of A and H to the vector bosons are given by
(\ref{EQU:psuver}) and (\ref{EQU:stanguv}), respectively.
In the massless fermion approximation, we find
\beq
\dd^8\Gamma_i
= C_i\Bigl[X_i +\sin(2 \chi_{1}) \sin(2 \chi_{2})Y_i \Bigr]
\dLips(m^2;q_1,q_2,q_3,q_4) , \qquad i=H,A,
\label{EQU:utgpkt}
\eeq
with $\dLips(m^2;q_1,q_2,q_3,q_4)$ denoting the Lorentz-invariant phase
space,
$m$ the Higgs mass,
and with the momentum correlations given by
\beqa
X_{\rm H}
& = &
(q_1\cdot q_3)(q_2\cdot q_4) +(q_1\cdot q_4)(q_2\cdot q_3), \\
Y_{\rm H}
& = &
(q_1\cdot q_3)(q_2\cdot q_4) -(q_1\cdot q_4)(q_2\cdot q_3), \\
X_{\rm A}
& = &
-2[(q_1\cdot q_2)(q_3\cdot q_4)]^2
-2[(q_1\cdot q_3)(q_2\cdot q_4) -(q_1\cdot q_4)(q_2\cdot q_3)]^2 \nn \\
&   &
+(q_1\cdot q_2)(q_3\cdot q_4)
\{[(q_1\cdot q_3)+(q_2\cdot q_4)]^2 +[(q_1\cdot q_4)+(q_2\cdot q_3)]^2\}, \\
Y_{\rm A}
& = &
(q_1\cdot q_2)(q_3\cdot q_4) \:
[(q_1-q_2)\cdot(q_3+q_4)][(q_3-q_4)\cdot(q_1+q_2)].
\label{EQU:eqYA}
\eeqa
The normalizations are given as
\beq
C_{\rm H}
 =
\frac{\sqrt{2}G_{\rm F}m_{\rm V}^4}{m}\:
D(s_1,s_2)
\qquad\mbox{\rm and} \qquad
C_{\rm A}
 =
\frac{g_{\rm eff}^2(s_1,s_2)}{4 m}\:
D(s_1,s_2) ,
\eeq
with
\beq
D(s_1,s_2) = \prod_{j=1}^{2}
\frac{g_j^2}{(s_j-m_{\rm V}^2)^2+m_{\rm V}^2\Gamma_{\rm V}^2}\:
N_j\:, \label{EQU:defc}
\eeq
and
\beq
s_1 \equiv (q_1+q_2)^2,   \qquad
s_2 \equiv (q_3+q_4)^2.   \label{EQU:def4}
\eeq
Here, $N_j$ is a colour factor, which is three for quarks,
and one for leptons.
Finally, $m_{\rm V}$ and $\Gamma_{\rm V}$ denote the mass and total width of
the relevant vector boson, respectively.

We first turn to a discussion of angular correlations.
The relative orientation of the two planes is defined by the angle
$\phi$,
\beq
\cos \phi = \frac{\left(\vec{q}_{1} \times \vec{q}_{2}\right)
\cdot \left(\vec{q}_{3} \times \vec{q}_{4}\right)}{|\vec{q}_{1}
\times \vec{q}_{2}| |\vec{q}_{3} \times \vec{q}_{4}|}.
\label{EQU:Dj4}
\eeq
We find
\beqa
\frac{\dd^3\Gamma_{\rm H} }{\dd\phi\: \dd s_1 \dd s_2}
& = &
\frac{8 \sqrt{2} G_{\rm F} m_{\rm V}^{4}}{9}
\frac{D(s_1,s_2)\sqrt{\lambda\left(m^2,s_1,s_2\right)}}
{(8 \pi)^6 m^3}
\Biggl[\lambda\left(m^2,s_1,s_2\right)+12 s_1 s_2 \label{EQU:Dk1} \\
&   & +
\left(\frac{3 \pi}{4}\right)^2
\sqrt{s_1 s_2}(m^2-s_1-s_2)
\sin(2 \chi_{1}) \sin(2 \chi_{2}) \cos \phi
+ 2s_1 s_2 \cos 2 \phi \Biggr], \nn \\
\frac{\dd^3\Gamma_{\rm A}}{\dd\phi\: \dd s_1 \dd s_2}
& = &
\frac{g_{\rm eff}^2\left(s_{1},s_{2}\right)}{9}
\frac{D(s_1,s_2)\left[\lambda(m^2,s_1,s_2)\right]^{3/2}}
{(8 \pi)^{6} m^3}\:
s_{1} s_{2} \: [4 - \cos 2 \phi],
\label{EQU:Dk2}
\eeqa
where
$\lambda\left(x,y,z\right)=
x^{2}+y^{2}+z^{2}-2\left(x y + x z + y z\right)$
is the usual two-body phase space function.
The term $Y_{\rm A}$ of eq.~(\ref{EQU:eqYA}) does not contribute in
eq.~(\ref{EQU:Dk2}).

In order to obtain a more inclusive distribution, we shall next integrate
over the invariant masses of the two pairs.
Thus,
\beq
\frac{\dd\Gamma}{\dd\phi}
= \int_{0}^{m^2}\dd s_1
\int_{0}^{\left(m -\sqrt{s_1}\right)^2}\dd s_2 \:
\frac{\dd^3\Gamma}{\dd\phi \: \dd s_1 \: \dd s_2}. \label{EQU:Dk3}
\eeq
Integrating over $0\le\phi< 2 \pi$ in the standard-model case, we
confirm the result quoted in ref.~\cite{Kniehl} for Higgs decay into four
fermions including finite-width effects.

We introduce the ratios
\beq
\mu = \left(\frac{m_{\rm V}}{m}\right)^2
\qquad\mbox{\rm and} \qquad
\gamma = \left(\frac{\Gamma_{\rm V}}{m}\right)^2,
\label{EQU:Dk6}
\eeq
and change variables according to $x_1=s_1/m^2$ and $x_2=s_2/m^2$.
This enables us to define the integrals
\beqa
F(m)
& \equiv &
\int_{0}^{1}
\frac{\dd x_{1}}{\left(x_{1}-\mu\right)^{2}+\mu \gamma} \nn \\
&        &
\times
\int_{0}^{\left(1-\sqrt{x_{1}}\right)^{2}} \dd x_{2} \:
\frac{\sqrt{\lambda\left(1,x_{1},x_{2}\right)}}
{\left(x_{2}-\mu\right)^{2}+\mu \gamma} \left[
\lambda\left(1,x_{1},x_{2}\right)+12 x_{1} x_{2} \right],
\label{EQU:Dk8}
\\
A(m)
& \equiv &
\int_{0}^{1}
\frac{\dd x_{1} \,\sqrt{x_1}}{\left(x_1-\mu\right)^2+\mu \gamma}
\int_{0}^{\left(1-\sqrt{x_1}\right)^2} \dd x_{2}\:
\frac{\sqrt{x_{2}} \sqrt{\lambda\left(1,x_{1},x_{2}\right)}}
{\left(x_{2}-\mu\right)^{2}+\mu \gamma} \left(1-x_{1}-x_{2}
\right),
\label{EQU:Dk9}
\\
B(m)
& \equiv &
\int_{0}^{1}
\frac{\dd x_{1}\, x_{1}}{\left(x_{1} -\mu\right)^{2}+\mu \gamma}
\int_{0}^{\left(1-\sqrt{x_{1}}\right)^{2}} \dd x_{2} \:
\frac{x_{2}\sqrt{\lambda\left(1,x_{1},x_{2}\right)}}
{\left(x_{2}-\mu\right)^{2}+\mu \gamma}.
\label{EQU:Dk10}
\eeqa
The distributions of eq.~(\ref{EQU:intro1}) then take the form
\beqa
\frac{2 \pi}{\Gamma_{\rm H}}\:\frac{\dd\Gamma_{\rm H}}{\dd\phi}
& = &
1 + \alpha\left(m\right)\sin(2 \chi_{1})\sin(2 \chi_{2}) \cos \phi
+ \beta\left(m\right)\cos 2 \phi ,
\label{EQU:Dl5} \\
\frac{2 \pi}{\Gamma_{\rm A}}\:\frac{\dd\Gamma_{\rm A}}{\dd\phi}
& = &
1 - \frac{1}{4} \cos 2 \phi ,
\label{EQU:Dl6}
\eeqa
where
\beq
\alpha(m)=\left(\frac{3 \pi}{4}\right)^2 \frac{A(m)}{F(m)},
\qquad
\beta(m)=2\,\frac{B(m)}{F(m)}.
\eeq
The functions $\alpha$ and $\beta$ depend on
the ratios of the masses of the vector bosons to the Higgs mass.
They are given in fig.\ref{plfun1}, for values of $m$ up to 1000~GeV.
In the narrow-width approximation these decay correlations
are identical to the ones reported in \cite{Nel} and our
analysis fully justifies this approximation.

However, our analysis is valid also below the threshold
for producing real vector bosons, $m < 2 \, m_{\rm V}$.
A representative set of angular distributions are given
in fig.\ref{plfun2} for the cases
${\rm H} \rightarrow (l^{+} \nu_{l}) (b \overline{c})$
and ${\rm H} \rightarrow (l^{+} l^{-}) (b \overline{b})$ for
$m=70, 150, 300$~GeV and $m=70, 300$~GeV, respectively.
(Of course, jet identification is required for this kind of analysis.)
With $\phi$ being defined as the angle between two {\it oriented} planes,
it can take on values $0\le\phi\le\pi$.
There is seen to be a clear difference between the CP-even
and the CP-odd cases.
We observe that the distribution (\ref{EQU:Dl5}) becomes
uncorrelated in the heavy Higgs limit, whereas the
distribution (\ref{EQU:Dl6}) is independent of the Higgs mass.

Let us now turn to a discussion of energy differences.
In order to project out the $Y_{\rm A}$-term of eq.~(\ref{EQU:eqYA}),
we multiply (\ref{EQU:utgpkt}) by the energy differences
$(\omega_{1}-\omega_{2})(\omega_{3}-\omega_{4})$ before
integrating over energies.
In analogy with eq.~(\ref{EQU:utgpkt}), we introduce
\beq
\dd^8\tilde{\Gamma}_i
= C_i\Bigl[\tilde{X}_i +\sin(2 \chi_{1}) \sin(2 \chi_{2})\tilde{Y}_i \Bigr]
\dLips(m^2;q_1,q_2,q_3,q_4) ,
\label{EQU:utgpktmrk}
\eeq
with
\beqa
\tilde{X}_i & = & X_i (\omega_{1}-\omega_{2})(\omega_{3}-\omega_{4}) ,
\label{EQU:mrk1} \\
\tilde{Y}_i & = & Y_i (\omega_{1}-\omega_{2})(\omega_{3}-\omega_{4}) ,
\label{EQU:mrk2}
\eeqa
for $i=$ H, A. The distribution takes the form
\beq
\frac{2 \pi}{\tilde{\Gamma}}
\:\frac{\dd\tilde{\Gamma}}{\dd\phi}
 =
1 + \frac{\kappa\left(m\right)}{\sin(2 \chi_{1})\sin(2 \chi_{2})}
\cos \phi,
\label{EQU:Dl7}
\eeq
in the CP-even case, whereas there is no correlation in the CP-odd case;
i.e. $\kappa\left(m\right)=0$ and in this case
the term $\tilde{X}_{\rm A}$ of eq.~(\ref{EQU:mrk1}) does not contribute.
Here,
\beq
\kappa(m) =
\frac{1}{2} \left(\frac{3 \pi}{16}\right)^2 \frac{K(m)}{J(m)},
\label{EQU:Dkappa}
\eeq
with
\beqa
J(m)
& \equiv &
\int_{0}^{1}
\frac{\dd x_{1}\,x_{1}}{\left(x_{1} -\mu\right)^{2}+\mu \gamma}
\int_{0}^{\left(1-\sqrt{x_{1}}\right)^{2}} \dd x_{2} \:
\frac{x_{2} \: \left[\lambda\left(1,x_{1},x_{2}\right)\right]^{3/2}}
{\left(x_{2}-\mu\right)^{2}+\mu \gamma},
\\ \label{EQU:Dk11}
K(m)
& \equiv &
\int_{0}^{1}
\frac{\dd x_{1} \,\sqrt{x_1}}{\left(x_1-\mu_1\right)^2+\mu_1 \gamma_1}
\nn \\
&        &
\times
\int_{0}^{\left(1-\sqrt{x_1}\right)^2} \dd x_{2}
\frac{\sqrt{x_{2}} \: \left[\lambda\left(1,x_{1},x_{2}\right)\right]^{3/2}}
{\left(x_{2}-\mu\right)^{2}+\mu \gamma} \left(1-x_{1}-x_{2}
\right).
\label{EQU:Dk12}
\eeqa
The function $\kappa$ is given in fig.\ref{plfun3}, for values of $m$ up to
1000~GeV.
In the narrow-width approximation
\beq
\kappa(m)=\frac{1}{16} \: \frac{\alpha(m)}{\beta(m)}.
\label{EQU:relasjon}
\eeq
The correlation is significant for any Higgs mass, and in particular
$\kappa(m) \propto m^{2}$ in the heavy-Higgs limit.
A representative set of distributions
(\ref{EQU:Dl7}) are given in fig.\ref{plfun4}
in the cases ${\rm H} \rightarrow (l^{+} \nu_{l})
(b \overline{c})$ and ${\rm H} \rightarrow (l^{+} l^{-})
(b \overline{b})$ for $m=70, 300, 500$~GeV and $m=70, 300$~GeV,
respectively.
We see that the energy-weighted distribution
is a much more sensitive probe for CP determination than the
purely angular distribution of eq.~(\ref{EQU:intro1}).
In this latter case, we compare an uncorrelated
distribution with a strongly correlated one.
Moreover, for ${\rm V}={\rm Z}$ the $\sin 2\chi$-factors
provide an enhancement in the energy-weighted correlations.



This research has been supported by the Research Council of Norway.

%
\renewcommand{\arraystretch}{0.6}
\begin{table}[p]
\vspace{0.5cm}
\begin{center}
\begin{tabular}{|c|c|c|}    \hline
${\rm V}$                                  & $f$
      & $\sin 2 \chi$ \\
  \hline
${\rm W}$                                  & any
      & 1 \\
${\rm Z}$                                  & e, $\mu$, $\tau$
      & 0.1393 \\
${\rm Z}$                                  & u, c, t
      & 0.6641 \\
${\rm Z}$                                  & d, s, b
      & 0.9349 \\
  \hline
\end{tabular}
\end{center}
\caption{The factors $\sin 2 \chi$ that give the relative importance of the
axial couplings. \label{TABLE:
sinkji}}
\end{table}
%
\renewcommand{\arraystretch}{1.5}
\clearpage
\centerline{\bf Figure captions}

\vskip 15pt
\def\fig#1#2{\hangindent=.65truein \noindent \hbox to .65truein{Fig.\ #1.
\hfil}#2\vskip 2pt}

\fig1{The coefficients $\alpha$ and $\beta$ of the angular correlations in
eq.~(\ref{EQU:Dl5}), for a CP-even Higgs of mass $m$.}

\fig2{Characteristic angular distributions of the planes defined by two
Dalitz pairs for CP-even Higgs particles decaying via two ${\rm W}$'s and two
${\rm Z}$'s, compared with the corresponding distribution for a CP-odd Higgs
particle (denoted ${\rm A}$). Different Higgs masses are considered in the
CP-even case.}

\fig3{The coefficient $\kappa$ of the energy-weighted angular correlation
of eq.~(\ref{EQU:Dl7}), for a CP-even Higgs of mass $m$.
For a CP-odd Higgs, $\kappa=0$.}

\fig4{Characteristic energy-weighted distributions for CP-even Higgs
particles decaying via two ${\rm W}$'s and two
${\rm Z}$'s, compared with the corresponding distribution for a CP-odd Higgs
particle (${\rm A}$).}

\clearpage

\begin{figure}
\refstepcounter{figure}
\label{plfun1}
\begin{center}

\vspace{0mm}

Figure~\thefigure
\end{center}
\end{figure}
\clearpage

\begin{figure}
\refstepcounter{figure}
\label{plfun2}
\begin{center}

\vspace{0mm}

Figure~\thefigure
\end{center}
\end{figure}
\clearpage

\begin{figure}
\refstepcounter{figure}
\label{plfun3}
\begin{center}

\vspace{0mm}

Figure~\thefigure
\end{center}
\end{figure}
\clearpage

\begin{figure}
\refstepcounter{figure}
\label{plfun4}
\begin{center}

\vspace{0mm}

Figure~\thefigure
\end{center}
\end{figure}

\end{document}